\documentstyle[aps,multicol,epsf]{revtex}  
\begin{document}
\twocolumn[\hsize\textwidth\columnwidth\hsize\csname 
@twocolumnfalse\endcsname                            
\title{
Stability of the trapped nonconservative Gross-Pitaevskii
equation with attractive two-body interaction
}
\author{
Victo S. Filho$^{1}$,
T. Frederico$^{2}$,
Arnaldo Gammal$^{1,3}$, and 
Lauro Tomio$^{1}$
}
\address{
$^{1}$ Instituto de F\'{\i}sica Te\'{o}rica,
Universidade Estadual Paulista, 01405-900, S\~{a}o Paulo, Brazil\\
$^{2}$ Departamento de F\'{\i }sica, Instituto Tecnol\'{o}gico da
Aeron\'{a}utica, CTA, 12228-900, S\~{a}o Jos\'{e} dos Campos, 
Brazil\\ 
$^{3}$Instituto de F\'\i sica, Universidade de S\~ao Paulo,
05315-970, S\~ao Paulo, Brazil
}
\maketitle
\begin{abstract}
The dynamics of a nonconservative Gross-Pitaevskii equation for 
trapped atomic systems with attractive two-body interaction is 
numerically investigated, considering wide variations of the
nonconservative parameters, related to atomic feeding and dissipation. 
We study the possible limitations of the mean field description for an 
atomic condensate with attractive two-body interaction, by defining the 
parameter regions where stable or unstable formation can be found. 
The present study is useful and timely considering the possibility
of large variations of attractive two-body scattering lengths, which may 
be feasible in recent experiments.
\end{abstract}
\vskip 0.5cm ]                              
\section{Introduction}
The stability of the condensed state is governed by the nature of the 
effective atom-atom interaction: the two-body pseudopotential is 
repulsive for a positive $s-$wave atom-atom scattering length and it is 
attractive for a negative scattering length~\cite{huang}. The ultra-cold 
trapped atoms with repulsive two-body interaction undergoes a 
phase-transition to a stable condensed state, in several cases found 
experimentally, as for $^{87}$Rb~\cite{and95}, $^{23}$Na~\cite{mew95} and 
$^1$H~\cite{hidr}. However, a condensed state of atoms with negative 
$s-$wave atom-atom scattering length (as in case of $^{7}$Li~\cite{brad97}) 
would be unstable, unless the number of atoms $N$ is small enough such 
that the stabilizing force provided by the
zero-point motion and the harmonic trap 
overcomes the attractive interaction, as found on theoretical 
grounds~\cite{rup95,baym96}. Particularly, in the case of $^{7}$Li 
gas~\cite{brad97}, for which the $s-$wave scattering length is 
$a=-14.5\pm 0.4$ \AA, it was experimentally observed that the number of 
allowed atoms in the Bose condensed state was limited to a maximum value 
between 650 and 1300, a result consistent with the mean-field 
prediction~\cite{rup95}, where the term proportional to the two-body 
scattering length (negative) dominates the nonlinear part of the 
interaction. 

More recently, the maximum critical number of atoms for Bose-Einstein 
condensates with two-body attractive interactions have been deeply 
investigated by the JILA group, considering experiments with 
$^{85}$Rb~\cite{JILA}. They have considered a wide tunning of the 
scattering length, $a$, from negative to positive, by means of Feshbach 
resonance~\cite{Fesh,avaria}, and observed that the system collapses 
for a number of atoms smaller than the theoretically predicted number.
Their experimental results, when compared with theoretical predictions for
spherical traps, show a deviation of up to 20\% in the critical 
number. More precisely, it was shown in Ref.~\cite{brief} that part of 
this discrepancy is due to the non-spherical symmetry that was considered
in Ref.\cite{JILA}. Such a deviation can also be an indication of
higher order non-linear effects that one should take into account
into the mean-field description. 
In Ref.~\cite{GFTC}, it was considered the
possibility of a real and positive quintic term, due to three-body
effects, in the Gross-Pitaevskii formalism. A negative quintic term would 
favor the collapse of the system for a smaller critical number of atoms, 
as verified in the JILA's experiments.
However, the real significance of a quintic term in the formalism is still 
an open question.

Our main motivation in the present work is to analyze the dynamics 
represented by an extension of the mean-field or Gross-Pitaevskii 
approximation, with non-conservative imaginary terms that are added
to the real part of the effective interaction, the two-body nonlinear term 
with a spherically symmetric harmonic trap. 
For the imaginary part, the interaction is a combination of a linear term, 
related to atomic
feeding, and a quintic term, due to three-body recombination, that
is responsible for the atomic dissipation. This is an approximation 
that is commonly used to study the properties of Bose-Einstein condensed 
systems. We consider a wide variation of the nonconservative parameters,
in particular motivated by the actual realistic scenario, that already 
exists, of altering experimentally the two-body scattering 
length~\cite{avaria}. 
As it will be clear in the following, this possibility will lead 
effectively to a modification of the dissipation parameter. 
By changing the absolute value of the scattering length, from zero to 
very large absolute values, one can change in an essential way the 
behavior of the mean-field description. 
As it will be shown from the present numerical approach, the
results for the dynamical observables of the system can be
very stable (solitonic-type) or very unstable (chaotic-type);
the characteristic of the results will depend essentially on
the ratio between the nonconservative parameters related to 
the atomic feeding and dissipation.

In the next section, we review the formalism. The main results are
presented in section III, followed by our 
conclusions in section IV.

\section{Mean field approximation}
The mean field approximation has shown to be appropriate to describe 
atomic Bose-Einstein condensation of a dilute gas of atoms confined by a 
magnetic trap \cite{Dalfovo}. 
In the case of positive scattering length, we have a very good agreement
with experimental data, as the thermal cloud is practically absent (removed by
cooling evaporation) and almost all the particles are in the condensed state. 
In this case, the mean field approximation results 
in a nonlinear Schr\"odinger equation (NLSE) known as Gross-Pitaevskii 
equation (GPE). If we have $N$ particles trapped in a spherical 
harmonic potential, this equation is given by
\begin{eqnarray}
i\hbar \frac{\partial\Psi}{\partial t}=\left(-\frac{\hbar^{2}}{2m}
\vec{\nabla}^{2}+\frac{1}{2}m\omega^{2}r^{2}+\frac{4\pi\hbar^{2}a}{m}
\mid \Psi \mid^{2}\right)\Psi\,,
\label{gpe}
\end{eqnarray}                             
where $\Psi\equiv\Psi(\vec{r},t)$, the wave-function of the condensate, 
is normalized to the number $N$, $m$ is the mass of a single 
atom, $\omega$ is the angular frequency of the trap, and $a$ is the 
two-body scattering  length. 

In this work, we have concentrated our study in the interesting 
dynamics that  occurs when the scattering length is negative ($a=-|a|$).
In this case, it is well known that the system is unstable without the 
harmonic trap; and the trapped system has a critical limit $N_c$ in the number 
of condensed atoms. The mean-field approximation has also shown to be 
reliable in determining the critical number of particles and even 
collapse cycles in the condensate~\cite{brad97,KMS,edw95}.
Actually, systems with attractive two-body interaction 
are being intensively investigated experimentally~\cite{JILA}, by using 
the so-called Feshbach resonance~\cite{Fesh,avaria}. The scattering 
length can be tuned over a large range by adjusting an external magnetic 
field (for more details, see Ref.~\cite{Timm}). 
Here, we are interested in the dynamics of a realistic system, where we 
add two non-conservative terms:
one (linear) related to the atomic feeding from the non-equilibrium thermal 
cloud; and another, dissipative due to three-body recombination 
processes (quintic). 
It is true that other dissipative terms can also 
be relevant for an arbitrary trapped atomic system, as a cubic one, 
that can be related with dipolar relaxation or with an 
imaginary part of the two-body scattering length. However, in order to 
simplify the study and better analyze the results, we restrict our
considerations to the case that we have just one parameter related with 
the feeding and another related with dissipation. We have considered only 
the three-body recombination parameter for dissipation also motivated 
by the observation that, for higher densities, this term dominates the 
two-body loss~\cite{ueda}.
So, for the generalization of Eq.~(\ref{gpe}), we add the 
imaginary terms in the interaction, such that
\begin{eqnarray}
{\rm i}\hbar \frac{\partial\Psi}{\partial t}=&&
-\frac{\hbar^{2}}{2m}
\vec{\nabla}^{2}\Psi+\frac{1}{2}m\omega^{2}r^{2}\Psi 
+\frac{4\pi\hbar^{2}a}{m}\mid \Psi\mid^{2}\Psi 
\nonumber\\
&&+ {\rm i}{G_\gamma}\Psi 
-{\rm i}G_\xi\mid\Psi\mid^{4}
\Psi\,,
\label{ggpe}
\end{eqnarray}                                  
where $G_{\xi}$ is the dissipation parameter, due to three-body 
collisions, and $G_\gamma$ is a parameter related to the feeding of the 
condensate from the thermal cloud. The Eq.~(\ref{ggpe}) was first suggested in 
Ref.~\cite{KMS} to simulate the condensation of $^7$Li. 

In order to recognize easily the physical scales in Eqs.~(\ref{gpe}) 
and (\ref{ggpe}), it is convenient to work with dimensionless units. By making 
the transformations 
\begin{eqnarray}
&&\vec{r}\equiv \sqrt{\frac{\hbar}{2 m\omega}}\vec{x}, \;\;\;\;\;
t\equiv \frac{\tau}{\omega} ,\label{rt}\\
&&G_\gamma\equiv\frac{\gamma}{2}{\hbar\omega}
,\;\;\;\;\;
G_\xi\equiv{2\xi}
\left(\frac{4\pi|a|\hbar}{m\omega}\right)^2 {\hbar\omega}\,\,\,\,\,\,
{\textnormal{and}}
\label{gs}\\
&&\Phi\equiv\Phi(x,\tau)\equiv\sqrt{8\pi|a|}|\vec r|\Psi(\vec{r},t),\label{wf}
\end{eqnarray}
we obtain the radial dimensionless $s-$wave equation:
\begin{eqnarray}
{\rm i}\frac{\partial\Phi}{\partial\tau}
&=&
\left[-\frac{d^{2}}{dx^{2}}+\frac{x^2}{4}-\frac{|\Phi|^{2}} {x^{2}}
-2{\rm i}\xi \frac{|\Phi|^{4}}{x^{4}}+{\rm i}\frac{\gamma}{2}\right] \Phi\,.
\label{tschd} 
\end{eqnarray}     
As $\Psi(\vec{r},t)$ is normalized to the number of atoms $N(t)$
in Eq.~(\ref{ggpe}), the corresponding time-dependent normalization
of  $\Phi(x,\tau)$ is given by the reduced number $n(\tau)$:
\begin{equation}
\int_0^\infty dx |\Phi(x,\tau)|^2 = n(\tau)
\equiv 2N(t)|a|\sqrt{\frac{2m\omega}{\hbar}}\,.
\label{norm}
\end{equation}
The nonconservative GPE (\ref{tschd}) is valid in the mean-field 
approximation of the quantum many-body problem of a dilute gas, when 
the average inter-particle distances are much larger than the absolute 
value of the scattering length; and also when the wave-lengths are much 
larger than the average inter-particle distance. The nonconservative 
terms are important when the condensate oscillates, fed by 
the thermal cloud, while losing atoms due to three-body inelastic 
collisions, which happen mainly in the high density regions.   

In order to verify the stability and the time evolution of the 
condensate, as observed in Refs.~\cite{FGFT}, two possible relevant 
observables are the number of particles normalized by the critical number 
of atoms of the static case ($N(t)/N_{c}$) and the mean square radius
(msr), 
\begin{eqnarray}
\langle r^2(t) \rangle &=& 
\left(\frac{\hbar}{2m\omega}\right)
{\frac{1}{n(\tau)} \int_0^\infty dx\; x^2|\Phi(x,\tau)|^2 
}\nonumber\\
&\equiv& \left(\frac{\hbar}{2m\omega}\right)
{\langle \ x^2(\tau) \ \rangle}
\equiv \left(\frac{\hbar}{2m\omega}\right) X^2\, ,
\label{msr}
\end{eqnarray}
where $X\equiv \sqrt{\langle \ x^2(\tau) \ \rangle}$ is the
dimensionless root mean square radius (rmsr).
In our analysis of stability, we calculate the time evolution of these 
quantities. We explore several combinations of the dimensionless 
nonconservative parameters $\xi$ and $\gamma$.
We first consider the case in which the atomic feeding is absent or when its 
parameter is smaller than the atomic dissipation parameter. Next, we explore 
variations of both parameters of about five orders of magnitude, from 
$10^{-5}$ to $10^{-1}$. This wide spectrum includes 
the parameters considered by Kagan {\it et al.}~\cite{KMS}, as well as 
other combinations that can be considered more realistic due to recent 
experimental results~\cite{GSFH}. 

Actually, the relevance of a wider relative variation of the 
nonconservative parameters $\gamma$ and $\xi$, presented in 
Eq.~(\ref{tschd}), can be better appreciated in face of the 
experimental possibilities that exist to alter the two-body scattering
length~\cite{avaria}. As one should note from Eq.~(\ref{gs}), any 
variation of the scattering length will also affect the effective 
dissipation parameter $\xi$ and, consequently, its relation with the 
feeding parameter $\gamma$. This implies that, by changing the value
of the scattering length, from positive to negative, and from zero to very
large absolute values, one can change in an essential way the 
behavior of the mean-field description. In the present work, we are 
concerned with negative two-body scattering length, where the collapsing
behavior of the Eq.~(\ref{tschd}) shows a very interesting dynamical 
structure. Even considering the possible limitations on the validity of the 
mean-field approach after the first collapse (in cases of parametrization
where it can occur), it is worthwhile to verify 
experimentally the behavior of a system in such a situation, by
varying $|a|$. At least, one can verify how far the theoretical description
can be qualitatively acceptable.

As already verified for systems with attractive interaction, as the 
$^7$Li, it has been possible, via the mean-field approach, to describe 
properties like the critical number of atoms in the condensate and growth and 
collapse cycles \cite{brad97,KMS,edw95}; besides, in the long time 
evolution, 
for certain sets of parameters, the calculations have also shown the 
presence of strong instabilities of the condensate, with 
signals of spatio-temporal chaotic behavior. 

In order to characterize a chaotic behavior, it is necessary
to show that the largest Lyapunov exponent related with the solutions
of the equation is positive. We follow the criterion used by Deissler and 
Kaneko \cite{deissler} to characterize spatio-temporal chaos. This criterion 
prescribes that the largest Lyapunov exponent for the system, in an arbitrary 
time interval, is obtained by plotting the logarithm of a function
$\zeta$, which is defined by
\begin{eqnarray}
\zeta(\tau)&\equiv&\left(\int_{0}^{\infty}|\delta\Phi(x,\tau)|^{2}dx\right)^
{1/2}
\label{zeta} .
\end{eqnarray}
$\delta\Phi(x,\tau)$ will give us the separation between two
nearby trajectories; it is obtained in the following form:
we numerically evolve in time an initial $\Phi_0(x)$, obtaining 
$\Phi(x,\tau)$. Independently, we evolve $\Phi_0(x) + \epsilon(x)$, 
and get $\Phi'(x,\tau)$, where $\epsilon(x)$ is a very small 
random perturbation.
$\delta\Phi(x,\tau)$ is given by 
$\Phi'(x,\tau) - \Phi(x,\tau)$. 
The chaotic behavior is characterized by a positive slope of 
$\ln\zeta(\tau)$, which gives the largest Lyapunov 
exponent~\cite{deissler}.

\section{Numerical Results}

In the next, we present the most significant results that
characterize the time evolution of the normalized number of particles 
( $N(t)/N_{c}$ ), the dimensionless mean square radius 
$\langle x^2\rangle$, and, in order to characterize the stability of the 
system, the function related to the largest Lyapunov exponent.
Further, we present a representative case of the 
phase-space for the root mean square radius. 
We have studied a wide region of parameters $\gamma$ and $\xi$, covering 
about five orders of magnitude, from $10^{-5}$ to $10^{-1}$, including 
the case with no feeding ($\gamma=0$). 

In order to have a clear and useful map of the regions where one should 
expect stable results, as well as regions with instabilities or chaos,
we summarize the present numerical results in a diagrammatic picture that 
relates these two nonconservative parameters. 
In general, it is expected that the system is more stable when the 
parameter related to the feeding of atoms ($\gamma$) from the  thermal 
cloud is significantly smaller than the parameter related to the 
dissipation ($\xi$). However, it is interesting to find out the 
region of parameters where this transition (from stable to unstable
results) occurs. Analysis of experimental results can provide a test to 
the present mean-field description in case of negative two-body 
interaction. 
As previously observed, we are considering dimensionless observables and 
parameters. For any realistic comparison with experimental parameters, one 
should convert $\gamma$ and $\xi$ to the parameters $G_\gamma$ 
and $G_\xi$, as given in Eq.~(\ref{gs}). 

The numerical solutions of Eq.~(\ref{tschd}) were obtained by applying 
the semi-implicit Crank-Nicolson algorithm for nonlinear problems, as 
implemented in Ref.~\cite{FGFT}. This method is stable and, therefore, very
convenient and reliable to treat time-dependent non-linear partial
differential equations. The initial 
condition for the number of atoms $N$ in the condensate was such that 
$N(0)/N_{c} = n(0)/n_{c} = 0.75$.
The evolution of the observables have been extended up to $\tau=\omega t=500$. 

In general, as expected, the smaller is the dissipation parameter, the 
longer is the life of the condensate. The mean square radius presents an
oscillatory behavior while one increases $\xi$. 
One observes that, in the regime of small feeding ($\gamma\le 10^{-4}$), the 
extended Lyapunov presents no positive slope. 
For larger values of $\gamma$, from $\sim 10^{-3}$ and $10^{-2}$, we 
have studied a few cases where the interplay between the nonconservative
behaviors are significant.

\begin{figure}
 \setlength{\epsfxsize}{1.0\hsize} 
\centerline{\epsfbox{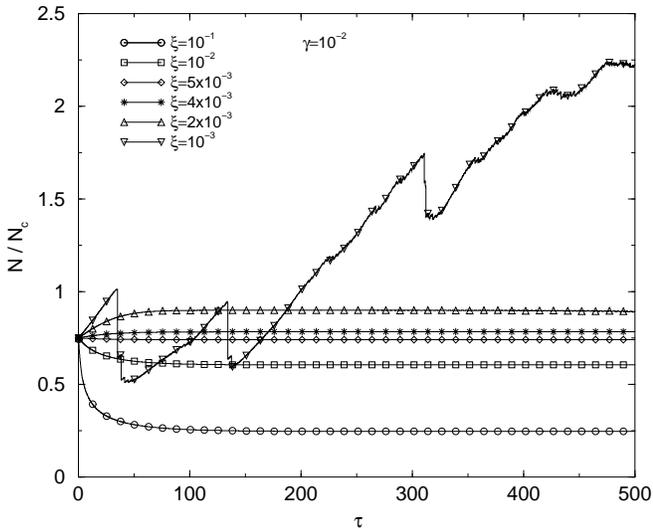}}
\caption[dummy0]{Time evolution of the number of condensed atoms $N$,
relative to the critical number $N_c$, for a set of values of the
dissipative parameter $\xi$ (as shown inside the frame), with the feeding
parameter $\gamma=10^{-2}$. 
All the quantities are in dimensionless units, as given in 
Eqs.~(\ref{rt})-(\ref{gs}).} 
 \label{FIG1}
 \end{figure}

\begin{figure}
 \setlength{\epsfxsize}{1.0\hsize} 
\centerline{\epsfbox{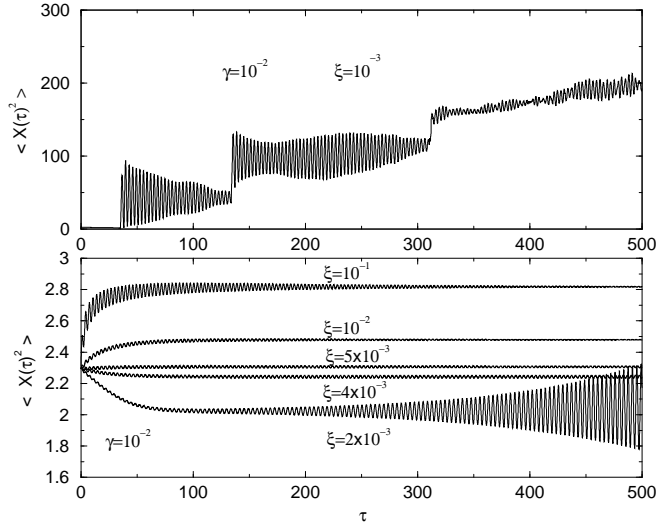}}
\caption[dummy0]{Time evolution of the dimensionless 
mean\--square\--radius, 
$\langle x(\tau)^2\rangle$, for the feeding parameter $\gamma=10^{-2}$. 
The results are given for a set of values of the dissipative parameter 
$\xi$, in the lower frame (shown inside). A specific case, for 
$\xi=10^{-3}$ (much smaller than $\gamma$), is isolated in the upper 
frame, where one can observe the behavior of $\langle x(\tau)^2\rangle$ after the 
collapse. All the quantities are in dimensionless 
units, as given in Eqs.~(\ref{rt})-(\ref{gs}).} 
 \label{FIG2}
 \end{figure}

In Fig.~\ref{FIG1}, we show the dynamical behavior of the number of atoms
for $\gamma=10^{-2}$ and several values of $\xi$; and, in the
Fig.~\ref{FIG2}, the corresponding time evolution of 
$\langle x(\tau)^2\rangle$.
We realize an interesting behavior, that occurs when the dissipation is larger 
than the feeding process: there are solutions of stability or dynamical 
equilibrium between both nonconservative processes. This phenomenon was
already discussed in Ref.~\cite{autosoliton}, for a few values of the 
dissipation and feeding parameters, using the time dependent variational 
approach and also the Crank-Nicolson method. 
In the present work, we observe a wide region of parameters where it is 
possible the formation of
autosolitons~\cite{autosoliton}. However, when the feeding process is
much larger than the dissipation, of about one or more orders of
magnitude, we can also observe chaotic behaviors. 
See, for example, the case with $\xi=10^{-3}$. 

The time evolution of the number of particles, represented in 
Fig.~\ref{FIG1}, shows a collapse for $\tau\approx$30, followed by several
other collapses, with the number of particles going above the critical
limit $N_c$. 
So, after a sequence of collapses, the critical limit for the number of
particles is no more followed, as already shown in Ref.~\cite{FGFT}.

The corresponding time evolution of $\langle x(\tau)^2\rangle$
is shown in the upper frame of Fig.~\ref{FIG2}. We observe that, 
following each collapse, after the shrinking of the system, the radius is
multiplied by a large factor, with indication of being populated by
radial excited states.
In the lower frame of Fig.~\ref{FIG2}, we can observe the corresponding
transition from the stable region (where the system finds the
equilibrium at a fixed value of the radius, corresponding to autosoliton
formation) to the unstable region. As shown, the instability starts to
occur when $\xi=2\times 10^{-3}$, and it can be developed to
a spatio-temporal chaos.  
The chaotic behavior can be verified through the 
Deissler and Kaneko criterion~\cite{deissler}.

\begin{figure}
 \setlength{\epsfxsize}{1.0\hsize} 
\centerline{\epsfbox{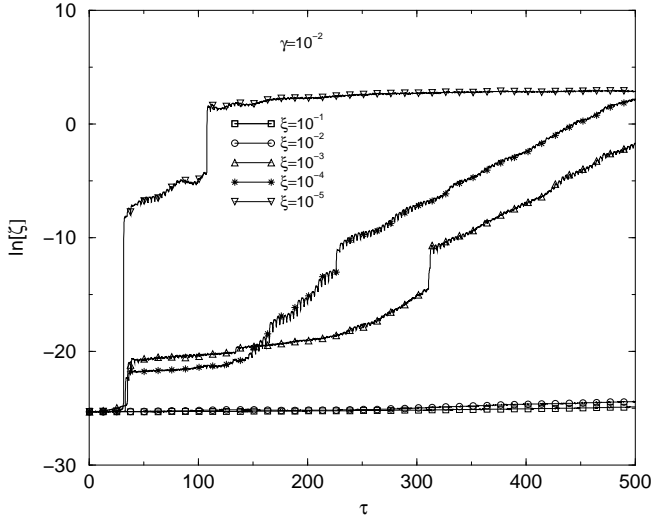}}
\caption[dummy0]{Time evolution of  ln$\zeta$, related to the
separation between two nearby trajectories [See Eq.~(\ref{zeta})], 
for $\gamma=10^{-2}$ and a set of values of $\xi$ indicated inside the 
figure. All the quantities are in dimensionless 
units, as given in Eqs.~(\ref{rt})-(\ref{gs}).} 
 \label{FIG3}
 \end{figure}

In Fig.~\ref{FIG3} we illustrate the application of the Deissler 
and Kaneko criterion to the system given by Eq.~(\ref{tschd}), for a fixed
value of the feeding parameter $\gamma=0.01$, and a set of values of
the dissipation parameter $\xi$. It was plotted the time
evolution of the function ln$(\zeta)$, where $\zeta$ is given by
Eq.~(\ref{zeta}), following the prescription given
in Ref.~\cite{deissler} to obtain the largest Lyapunov exponents for the
system. Within this prescription, the system becomes chaotic when
ln$(\zeta)$ has a positive slope. As shown in Fig.~\ref{FIG3}, this
clearly occurs, for example, when $\xi=10^{-4}$. In case of 
$\xi=10^{-5}$ we note a much faster increasing in ln$(\zeta)$, with
an observed saturation that happens due to the fact that such function
has reached the maximum separation between the trajectories. The
saturation properties is also verified when studying chaotic behaviors in
ordinary differential equations~\cite{strogatz}.
The plot of ln$(\zeta)$ corresponds to the same value of $\gamma$ 
($=10^{-2}$) used in Figs.~\ref{FIG1} and \ref{FIG2}.
As shown, a clear 
characterization of chaotic behaviors starts to occur only for values of 
the dissipation parameter $\xi$ much smaller than $\gamma$. In the cases 
presented in Fig.~\ref{FIG3}, for $\xi\le 10^{-3}$. 

\begin{figure}
 \setlength{\epsfxsize}{1.0\hsize} \centerline{\epsfbox{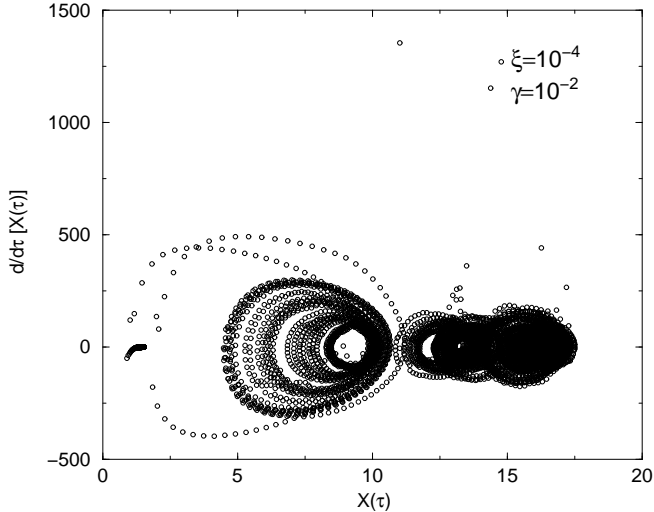}}
 \caption[dummy0]{ 
Phase-space for the root mean-square-radius, in dimensionless units
[$dX(\tau)/dt$ versus $X(\tau)$], considering a collapsing case that
leads to chaos. The dimensionless nonconservative parameters are 
$\xi=10^{-4}$ and $\gamma=10^{-2}$, and the time evolution was taken 
up to $t = \tau/\omega = 500/\omega$.}
 \label{FIG4} 
 \end{figure}

In Fig.~\ref{FIG4}, we present another significative illustration of 
chaotic behavior, through the phase-space behavior of the 
mean-square-radius, considering one case that was characterized as chaotic 
by using the Deissler-Kaneko criterion.
We have plotted in this figure the root mean-square-radius phase space
for the case with $\gamma=0.01$ and $\xi=10^{-4}$. 
The irregular behavior of the trajectories, observed in Fig.~\ref{FIG4}, 
with the classical strange attractors being observed,
clearly resembles chaos. This behavior is similar to the chaotic 
behavior observed in ordinary cases~\cite{strogatz}.

As a general remark that one can make from the presented results, we 
should note that, in order to observe unstable chaotic behaviors, the 
dissipation must be much smaller than the feeding parameter.

\begin{figure}
 \setlength{\epsfxsize}{1.0\hsize} \centerline{\epsfbox{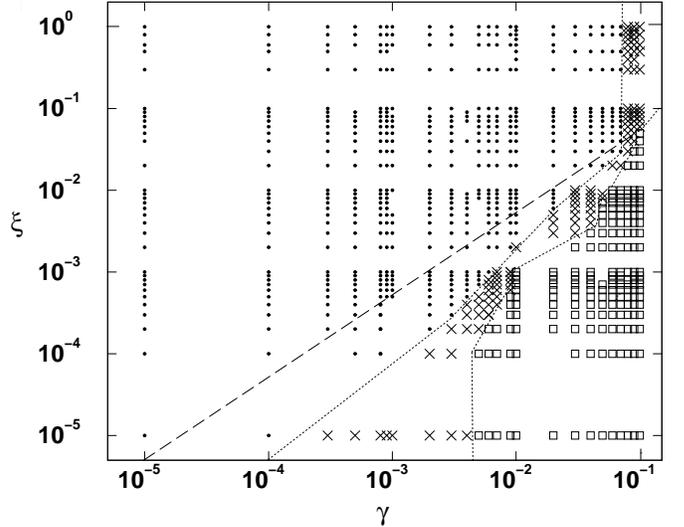}}
 \caption[dummy0]
{Diagram for stability, according to the criterion of Ref.~\cite{deissler} 
given by Eq.~(\ref{zeta}), with results for the equation (\ref{tschd}),
considering the dimensionless non-conservative parameters $\gamma$ and $\xi$. 
Between the unstable results, represented with $\times$ and squares, the 
chaotic ones are identified with squares. The stable results are 
represented by bullets. 
Two dotted guide-lines are splitting the regions.
The dashed line split the graph in two regions according to a
variational approach (see Ref.~\cite{autosoliton}); in the upper part the 
results are stable, in the lower, unstable.}
 \label{FIG5}
 \end{figure}

In a diagrammatic picture, given in Fig.~\ref{FIG5}, we resume our 
results. We show the relation between the two nonconservative parameters, 
$\xi$ and $\gamma$, in order to characterize the parametric regions where 
one should expect stability or instability in the solutions for the 
Eq.~(\ref{tschd}). The stable results of the Eq.~(\ref{tschd}) are 
represented by bullets; the nonstable results that clearly present 
positive slope for $\ln\zeta(\tau)$ (chaotic behavior) are represented by 
empty squares; with $\times$, we show other intermediate nonstable 
results, in which the characterization of chaotic behavior was not so 
clear, through the Deissler-Kaneko criterion. In this figure, in order
to observe the approximate consistency of the numerical results, we also 
include the variational analysis presented in Fig.1 of 
Ref.~\cite{autosoliton}, represented by the dashed-line. It is separating 
the stable region (upper part) from the unstable one (lower part).

We should note that, in section V of Ref.\cite{ehuang}, it was also 
considered the dynamics of growth and collapse, with non-conservative 
terms related to feeding ($\gamma_0$) and dissipation ($\gamma_1$ and
$\gamma_2$) in a specific example. For the dissipation they have also 
considered a term related to dipolar relaxation, given by $\gamma_2$.  
Here, in our systematic study of the regions of instability, we took 
into account previous experimental~\cite{JILA} observations that
the dominant process for the dissipation is the  three-body 
recombination. By comparing the parameters of Ref.~\cite{ehuang} with 
the parameters that we have used, and observing that our parameter $\xi$ 
should be related to both dissipation parameters used in 
Ref.~\cite{ehuang} ($\gamma=\gamma_0=2.6\times 10^{-3}$, 
$\xi\sim 10^{-5}$) one can verify from the results given in 
Fig.~\ref{FIG5} that the model of Ref.~\cite{ehuang} is inside the 
intermediate region, where the system is unstable, without a clear 
signature of chaos. 

\section{Conclusions}

In summary, we have studied the dynamics associated with the extended
nonconservative Gross-Pitaevskii equation for a wide region of the
dimensionless nonconservative parameters,
$\xi$ and $\gamma$, that, respectively, are related to atomic dissipation and
feeding in a trapped atomic condensed system. 
We consider systems with attractive two-body interaction in a spherically
symmetric harmonic trap.
In Fig.~\ref{FIG5}, we resume our results, by mapping the space of
$\gamma$ versus $\xi$, showing the regions of equilibrium and the
regions of instability, as well as the regions where we are able to 
characterize chaotic behaviors, using a criterion given in 
Ref.~\cite{deissler}. 
It was also confirmed that chaotic behaviors occur mainly when
$\gamma$ is big enough and $\gamma/\xi$ is large (at least, when 
$\gamma$ is one or two orders of magnitude larger than $\xi$). 
A wide variation of the nonconservative parameters was analyzed, 
in particular motivated by the actual realistic scenario, that already 
exists, of altering experimentally the two-body scattering 
length~\cite{avaria}. By changing the absolute value of the scattering length,
one can change in an essential way the behavior of the mean-field description. 

\section*{Acknowledgements}

We thank Funda\c c\~ao de Amparo \`a Pesquisa do Estado de S\~ao Paulo 
(FAPESP) for partial support. LT and TF also thank partial support from 
Conselho Nacional de Desenvolvimento Cient\'\i fico e Tecnol\'ogico (CNPq).

\end{document}